**Disruption of the prefrontal cortex improves implicit contextual memory-guided attention: combined behavioural and electrophysiological evidence**

Mario Rosero Pahi[1], Juliana Cavalli[1], Frauke Nees[1], Herta Flor[1*], Jamila Andoh[1*]

[1] Department of Cognitive and Clinical Neuroscience, Central Institute of Mental Health, Medical Faculty Mannheim, Heidelberg University, Square J5, 68159 Mannheim, Germany

*Joint last authors

Address correspondence to: jamila.andoh@zi-mannheim.de or herta.flor@zi-mannheim.de

**Abstract**

Many studies have shown that the dorsolateral prefrontal cortex (DLPFC) plays an important role in top-down cognitive control over intentional and deliberate behavior. However, recent studies have reported that DLPFC-mediated top-down control interferes with implicit forms of learning. Here we used continuous theta burst stimulation (cTBS) combined with electroencephalography (EEG) to investigate the causal role of DLPFC in implicit contextual memory-guided attention. We aimed to test whether transient disruption of the DLPFC would interfere with implicit learning performance and related electrical brain activity. We applied neuronavigation-guided cTBS to the DLPFC or to the vertex as a control region, prior to the performance of an implicit contextual learning task. We found that cTBS applied over the DLPFC improved significantly performance during implicit contextual learning. We also observed that beta-band (13-19 Hz) oscillatory power was reduced at fronto-central channels around 140 to 370 ms after visual stimulus onset in cTBS DLPFC compared with cTBS vertex. Taken together, our results provide evidence that the DLPFC-mediated top-down control interferes with contextual memory-guided attention and with beta-band oscillatory activity.

**Introduction**

Decades of research into the neurocognitive mechanisms of attention revealed that visual attention is fundamentally controlled by both exogenous (stimulus salience) and endogenous (internal goals and expectations) factors (Egeth and Yantis 1997; Shulman et al. 1999; Corbetta et al. 2000; Corbetta and Shulman 2002; Corbetta and Shulman 2011). Nevertheless, emerging evidence suggests that implicit and explicit memories also play an important role in guiding attention (Chun and Jiang 1998; Moores et al. 2003; Kristjánsson and Campana 2010; Stokes et al. 2012; Hutchinson and Turk-Browne 2012; Zhao et al. 2013). For instance, when subjects interact repeatedly with objects, which tend to co-vary with each other across space, contextual representations are spontaneously and unconsciously formed, and over time, those representations can be used to quickly direct the subjects' attention to the expected location of relevant stimuli. Thus, contextual memory-guided attention can be considered as a powerful mechanism to simplify the external world and generate predictions. Despite the potential importance of contextual memory-guided attention in cognition, its neural mechanisms and how these mechanisms are influenced by top-down processing have received little attention. Implicit contextual memory-guided attention has been largely studied using the contextual cueing paradigm (Chun and Jiang 1998). In this paradigm, subjects are instructed to search for a target item (e.g., the letter T) embedded in a spatial array of distractors items (e.g., a set of the letter L). When the target is embedded in a repeated configuration of arrays, participants' visual search has been shown to be faster compared to a novel configuration of arrays. In the first situation, the distractors items configure a visuo-spatial context that guides attention to the expected location of the target. Interestingly, participants do not explicitly recognize repeated configurations of arrays (Chun and Jiang 1998; Chun and Phelps 1999). Contrary to the supposition that contextual cueing is invariant among individuals, there is evidence that between 30% to 40% of the individuals do not show the contextual cueing effect (Lleras and von Mühlenen 2004; Schlagbauer et al. 2012). Different cognitive search strategies could be the reason for the performance variation in contextual cueing. Lleras and von Mühlenen (2004) found that when participants were instructed to search actively for the target, by deliberately shifting their attention, the contextual cueing effect was disrupted. By way of contrast, the contextual cueing effect was facilitated when participants adopt a passive search strategy. These findings suggest that voluntary cognitive control can interfere with implicit contextual memory-guided attention.

A substantial body of research indicates that cognitive control and the means to limit attention to goal-relevant information and suppress irrelevant distraction stem from a set of fronto-parietal regions, often called cognitive control network (Cole and Schneider 2007). Importantly, the dorsolateral prefrontal cortex (DLPFC), a region within the cognitive control network, has been specifically involved in exerting top-down cognitive control over intentional and deliberate behavior by suppressing automatic behaviors (Kübler et al. 2006; de Manzano et al. 2012). Therefore, it has been suggested that reducing cognitive control facilitates automatic behaviors and implicit forms of learning. Compelling evidence for the role of the DLPFC in implicit learning comes from a recent transcranial magnetic stimulation (TMS) study, which shows that inhibition of DLPFC function improved implicit recognition, suggesting that DLPFC-mediated explicit memory processes interfere with implicit recognition memory (Lee et al. 2013). Moreover, DLPFC inhibition and concomitant decrease in conscious self-monitoring and focused attention has been implicated in promoting implicit spontaneous associations (Limb and Braun 2008; Liu et al. 2012). In line with these findings, converging evidence from electroencephalographic (EEG) studies supports an important role of beta-band activity (BBA) in endogenous top-down cognitive control (Engel and Fries 2010). Interestingly, several studies indicate that task-related BBA predominates in paradigms that involve high cognitive control, whereas settings that require little or no cognitive control decrease BBA (Engel et al. 2001; Okazaki et al. 2008; Iversen et al. 2009; Engel and Fries 2010). Thus, reducing BBA could reflect low endogenous cognitive control.

Here, we addressed the question whether DLPFC-mediated top-down control interferes with implicit contextual memory-guided attention. To examine this question, we used continuous theta-burst stimulation (cTBS) to transiently disrupt the function of DLPFC and measured the resulting effects on behavioral and oscillatory responses in an implicit contextual memory task. We hypothesized that disruption of the DLPFC function by cTBS would improve implicit memory performance and decrease beta-band oscillatory activity.

**Material and methods**

*Participants*

Twenty-one healthy subjects (11 females; range 19-33 years) with no history of neurological or mental disorder participated in the study. Right-handedness was verified using the Edinburgh Handedness Inventory (Oldfield 1971). The Ethics Committee of the Medical Faculty Mannheim of the University of Heidelberg approved the study and written informed consent was obtained from all persons before participation.

*Experimental procedure*

Participants initially underwent magnetic resonance imaging (MRI) in order to acquire an anatomical scan, which was used to individually define the DLPFC target. Then, the participants were invited to the laboratory for two separate sessions, both consisting of concurrent TMS and EEG recordings (1-week interval between the two sessions). In the one session, the participants received cTBS over the DLPFC, and in the other session they received cTBS over the Vertex. The order of stimulation sites (i.e. DLPFC versus Vertex) was counterbalanced across participants. Immediately after cTBS administration, the participants completed a version of the contextual cueing task (Chun and Jiang, 1998). Finally, an explicit recognition test was carried out to assess awareness of the contextual displays.

*Task description: Laboratory experiment*

Repeated from above in the contextual cueing task, the subjects were encouraged to search a target item embedded in a spatial array of distractor items. The target was a T stimulus rotated 90 degrees to the right or to the left. The distractor stimuli were L-shapes presented randomly in one of four orientations (0°, 90°, 180°, 270°). Each display consisted of 12 items (a single target and 11 distractors) randomly positioned in an invisible 8 × 6 matrix (37.2° × 28.3°). For repeated displays, the target appeared in the same location within an invariant configuration of distractors across blocks. For new displays, the distractor configurations were newly generated in each block. Visual stimuli were presented on a gray background in a 17-inch Viewsonic VG710b LCD monitor.

Participants were seated 50 cm in front of the computer monitor. Each trial consisted of a fixation cross presented during 0.5 sec, followed by a display presentation (for a maximum of 3 sec). This was then followed by a new gray background screen with variable duration intertrial-interval (ITI; range 0.5 – 1 sec) preceding the next trial. During the display presentation, the subjects pressed one of the two buttons ("C" or "M") on a computer keyboard, corresponding to whether the bottom of the T was pointed to the right or to the left. An example of the trial sequence is shown in Fig. 1. Each subject performed 20 blocks (480 trials) of the search task with each block containing 24 intermixed trials of 12 new and 12 repeated displays. There were rest periods of 10 sec between blocks.

After block 20, the volunteers participated in an explicit recognition test in order to assess awareness of the contextual displays (Chun and Jiang, 1998). In this test, the 12 repeated displays used in the experimental session and 12 new displays were presented in random order. Participants were instructed to indicate whether each display presented was ''repeated'' or ''new'' by responding with one of two response buttons. The task was implemented in MATLAB (Math-Works, Natick, MA, USA) using the open-source Psychophysics toolbox 3 (Brainard 1997).

-------------------------------------------------------------

Insert Figure 1 around here

-------------------------------------------------------------

*Magnetic Resonance Imaging*

Magnetic resonance imaging was performed with a 3T MAGNETOM Trio whole body scanner (Siemens Medical Solutions, Erlangen, Germany) equipped with a standard 12-channel head coil. T1-weighted high resolution images were acquired with a magnetization-prepared rapid gradient echo (MPRAGE) sequence [TR 2300 ms, TE 2.98 ms, field of view 240 x 256 mm2, 160 sagittal slices, voxel size 1.0 x 1.0 x 1.0 mm$^3$, parallel imaging (GRAPPA) factor 2].

Determination of the TMS Sites from Magnetic Resonance Images

The location of the DLPFC target was chosen based on a previous study showing improvement in recognition accuracy after TMS applied over DLPFC using standard-space coordinates from the Montreal Neurological Institute (MNI) brain (i.e. -43, 35, 30; Lee et al., 2013). For each participant, the DLPFC target coordinates were transformed into the participant's native MRI space with the reversed MNI152 template-to-native transformation matrix employing FSL software. The individually-defined DLPFC coordinates were then used as a center to draw a 5mm-radius region-of-interest, which was the TMS target. In addition, we also applied TMS over the vertex to control for nonspecific effects of TMS, such as acoustic and somatosensory artifacts. The vertex was defined anatomically by the intersection between a sagittal line form the nasion to the inion and a coronal line from the tragus of both ears.

-----------------------------------------------------------
Insert Figure 2 around here
-----------------------------------------------------------

*Transcranial magnetic stimulation*

For each participant, we determined the resting motor threshold (MT), which is defined as the lowest stimulus intensity capable of eliciting a motor-evoked potential MEP > 50 µV in the relaxed abductor pollicis brevis (APB) in 5 of 10 consecutive trials at intervals > 5 s (Rothwell et al., 1999). TMS was delivered using a 75mm winding diameter figure-of-eight coil (MCF-B65) and a MagProX100 stimulator (MagProX100, MagVenture, Denmark).

In order to guide the coil over the DLPFC location for each participant, we used a neuronavigation procedure, which was performed using the Brainsight system (Rogue Research, Canada) equipped with an infrared camera for online subject tracking and coil positioning (Polaris Spectra, NDI, Canada).

After the motor threshold procedure, repetitive TMS was carried out using a continuous theta burst stimulation protocol (cTBS) consisting of three pulses at 50 Hz repeated at a frequency of 5 Hz. Thus, a total of 600 pulses were delivered over a period of 40 s preceding the start of the task (Huang et al. 2005). The stimulus intensities were set at 90% of each participant's resting motor threshold. TMS was applied with the coil held tangentially to the skull, with the handle pointing backward and laterally at a 45-degree angle away from the midline. TMS sessions were performed according to the published safety guidelines (Wassermann et al. 1996; Rossi et al. 2009).

*EEG acquisition*

The EEG was recorded from 30 scalp electrodes mounted in a BrainCap TMS (BrainProducts, Munich, Germany) at the following sites (M-22 Equidistant 32-Channel-Arrangement): Fp1, Fp2, F3, F4, C3, C4, P3, P4, O1, O2, F7, F8, T7, T8, P7, P8, Fz, Cz, Pz, Iz, FC1, FC2, CP1, CP2, FC5, FC6, CP5, CP6, TP9, TP10. Blinks and eye movements were monitored by using two pairs of bipolar electrodes placed approximately 1 cm lateral to the left and right external canthi (HEOG) and above and below the subject's right eye (VEOG). Electrode impedances were kept below 10 kΩ using conduction gel. AFz and FCz served as ground and reference electrode for online recording. Data were band-pass filtered from 0.016 to 250 Hz during recording and digitized with a sampling rate of 1000 Hz using BrainAmp amplifiers (BrainProducts, Munich, Germany).

*Analysis of behavioral data*

Accuracy and search reaction times (RTs) for correct trials were grouped into sets of four yielding five epochs in order to increase the power of the statistical analyses (Chun and Jiang 1998).

For accuracy and RTs, we carried out a three-way ANOVAs using contexts (new versus repeated), epochs (1-5) and cTBS sites (DLPFC versus vertex) as factors.

In addition, for RTs, we performed separate two-way repeated measures ANOVAs for each cTBS site (i.e., DLPFC, vertex), the factors were contexts (new versus repeated) and epochs (1-5).

The contextual cueing effect was calculated using the difference in RTs between new and repeated contexts (for each epoch). These values were then entered in a two-way repeated measures ANOVA with Context (new versus repeated) and epochs (1-5) as factors.

We also assessed interindividual variation in implicit contextual learning. For this purpose, the contextual cueing effect was calculated as the difference in RTs between new and repeated contexts collapsed across epochs 2-5.

Finally, for the explicit recognition test, a paired-sample t test on hits and false alarms was conducted.

For all analyses the alpha level was set to 0.05. Data are reported as mean ± SD unless stated otherwise.

*EEG data analysis*

Analysis of the EEG data was performed using EEGLAB 4.51 (Delorme and Makeig, 2004), Fieldtrip (Oostenveld et al. 2011) and custom scripts on the MATLAB 8.5 platform (Math-Works, Natick, MA, USA). Data were low-pass filtered at 120 Hz and high-pass filtered at 0.2 Hz (FIR filter) and downsampled to 500 Hz. For further analysis, epochs from -500 ms before to 800 ms after stimulus onset, were extracted. Trials containing non-stereotyped artefacts (e.g., cable movement, swallowing) and amplitudes of more than ±100 µV were removed. This led to the rejection of 19.3 % trials. Finally, to reduce the remaining artefacts (e.g., eyeblinks, horizontal eye movements, and electrocardiographic activity) an extended infomax independent component analysis (ICA) was applied on the data using a weight change $<10^{-6}$ as stop criterion (Bell and Sejnowski 1995), and components that reflected signal artefacts were removed from the EEG data. Artefact components were identified using visual inspection of their topography, power spectrum, and temporal dynamics. Spectral changes in total oscillatory activity (phase-locked and non-phase-locked to stimulus onset) were analyzed using the 'multitaper method' (Mitra and Pesaran 1999) based on discrete prolate spheroidal slepian sequences (Slepian 1978). Time-frequency representations (TFRs) were computed using a sliding window ($\Delta T= 250$ ms), applied in steps of 10 ms (-370 ms before to 370 ms after stimulus onset), with a frequency smoothing ($\Delta f= ±4$ Hz; 4–30 Hz), resulting in one taper applied to the sliding window. Then, the Fourier transforms of the tapered time windows were calculated. To estimate evoked oscillatory activity, we first averaged across trials for each condition before computing the spectral estimates. To estimate induced oscillatory activity, we first subtracted from each individual trial the evoked part before computing the spectral estimates. All the presented data represent baseline-normalized (-200 to -100 ms) relative changes in power:

$Pow(t,f)_{normalized} = (Pow(t, f) - Pow(f)_{baseline}) / Pow(f)_{baseline}$.

For statistical testing of differences in spectral power (repeated versus new context; cTBS DLPFC versus cTBS vertex), we used a cluster-based permutation test (Maris and Oostenveld 2007). In short, a simple dependent-samples t-test was calculated at each time point, frequency bin, and channel of the two experimental conditions. Then, all adjacent samples with t-values exceeding a preset significance level (5%) were grouped into clusters. The sum of t-values from the cluster with the maximum sum was then used as the test statistic. Next, a null distribution of cluster-level t-statistics was created by randomly shuffling the data across the two conditions using 1000 permutations. Cluster values greater or smaller than the 97.5th percentile ($P < 0.025$) were considered to be significant. To identify time windows and the frequency range of significant relative power differences, we first compared the spectral power between the repeated versus new context and then between cTBS DLPFC versus cTBS vertex within the 4 to 100 Hz frequency band and the post-stimulus interval from 0 to 370 ms, using a sliding-time window fieldtrip cluster statistic (Staudigl and Hanslmayr 2013; Waldhauser et al. 2016). Then, significant time-frequency windows were subjected to another cluster-based permutation test to identify significant locations. This resulted in a beta frequency range of interest between 13 to 19 Hz and 140 to 370 ms. To test for a possible interaction between context and cTBS sites, we first calculated the differences between (1) repeated and new contexts in cTBS DLPFC and (2) repeated and new contexts in cTBS vertex. Next, we contrasted these differences using a cluster-based permutation test.

**Results**

The TMS procedure was well tolerated and all participants completed the study.

*Behavioral performance*

Participants were highly accurate (98.3%). Accuracy did not differ between contexts (new versus repeated), ($F(1,20) = 0.19$, $P = 0.66$), epochs (1-5), ($F(4,80) = 1.07$, $P = 0.37$), or cTBS sites (DLPFC versus vertex), $F(1,20) = 0.04$, $P = 0.84$), and there was no significant interaction between contexts, epochs and cTBS sites ($F(4,80) = 2.00$, $P = 0.10$). In addition, no significant interaction was found between contexts and epochs, $F(4,80) = 0.88$, $P = 0.47$, contexts and cTBS sites, $F(1,20) = 3.10$, $P = 0.09$, or epochs and cTBS sites, $F(4,80) = 0.01$, $P = 0.99$.

For RTs, we found a significant three-way interaction between contexts (new versus repeated), epochs (1-5), and cTBS sites (DLPFC versus vertex), $F(4,80) = 3.48$, $P < 0.01$. To establish the source of these interactions, the data from each cTBS site were separately analyzed (Fig. 3A). For cTBS DLPFC, the main effect of contexts (new versus repeated) was significant, $F(1,20) = 75.28$, $P < 0.001$, indicating that search RTs were faster for repeated than for new contexts. There was a significant main effect of epochs (1-5), $F(4,80) = 43.94$, $P < 0.0001$, indicating that search RTs decreased during the task. A significant contexts by epochs interaction, $F(4,80) = 8.70$, $P < 0.0001$, showed that for cTBS DLPFC, search RTs for repeated contexts decreased more than for new contexts. Post hoc analyses using Bonferroni's correction ($P < 0.05$) indicated that for cTBS DLPFC, the mean RT was significantly faster for repeated than for new contexts only during epochs 2-5.

For cTBS vertex there was a significant main effect of contexts (new versus repeated), $F(1,20) = 18.81$, $P < 0.001$, indicating that search RTs were faster for repeated than for new contexts. There was a significant main effect of epochs (1-5), $F(4,80) = 52.51$, $P < 0.0001$, showing that search RTs decreased during the task. No significant interaction was found between contexts and epochs for cTBS Vertex, $F(4,80) = 0.55$, $P = 0.69$.

The analysis of magnitude of the contextual learning (contextual cueing effect) as a function of cTBS sites (DLPFC versus vertex) and epochs (1-5), Fig. 3B, revealed that there was a significant main effect of cTBS sites (DLPFC versus Vertex), $F(1,20) = 23.04$, $P < 0.001$, indicating that contextual learning performance differed between DLPFC and vertex cTBS. Additionally, a significant main effect of epochs (1-5) was found $F(4,80) = 7.89$, $P < 0.001$, showing that contextual learning increased during the task for both cTBS sites. There was a significant interaction between the effects of cTBS sites and epochs for implicit contextual learning performance, $F(4,80) = 3.48$, $P < 0.05$, indicating that cTBS DLPFC effect compared to cTBS vertex was different during the task.

Post hoc analyses using Bonferroni's correction (P < 0.05) revealed that the mean of the contextual learning performance for cTBS DLPFC was significantly higher than cTBS vertex only during epochs 2-5. Importantly, 100% and 72% of the participants showed the contextual cueing effect during cTBS DLPFC and cTBS Vertex, respectively (contextual cueing effect estimated across epochs 2-5).

*Recognition task*

Before the explicit recognition test, participants were asked if they noticed that some displays were repeated during the experiment. None of the participants reported noticing the repeated displays. For the explicit recognition test, participants recognized repeated contexts as repeated on 27% (hit rate) of the cTBS DLPFC trials and this did not differ from the percentage of new contexts misidentified as repeated contexts (false alarm), 27% (P = 0.28, Student's t-test). For the cTBS vertex condition, the hit rate was 27% and this did not differ from the false alarm rate, 30% (P = 0.57, Student's t-test). The hit rate and the false alarm rate did not significantly differ between the two cTBS conditions (P = 1.00, and P = 0.20, Student's t-test, respectively), Taken together, there was no evidence of explicit recognition of repeated contexts, in both conditions.

------------------------------------------------------------

Insert Figure 3 around here

------------------------------------------------------------

*Total oscillatory activity*

cTBS DLPFC significantly decreased task-related beta-band activity compared to cTBS vertex, in a frequency range of 13-19 Hz and a time window from 140 to 370 ms after stimulus onset (Fig. 4A right, B). This effect was restricted to the F3, F4 and P3 channels (cluster-corrected permutation test, P = 0.018, Fig. 3B). A time-frequency analysis of significant channels is shown in Fig. 3C. There was no significant difference in task-related oscillatory activity between repeated and new contexts. The sliding windows analysis did not show significant differences between the repeated and the new contexts in any time window or frequency range (Fig. 4A left). Finally, permutation testing did not show an interaction between cTBS sites and contexts in the beta-band (13-19 Hz, 140-370 ms, cluster-corrected permutation test, $P > 0.1$). This suggests that cTBS reduced the task-related beta-band oscillatory power independently of the contexts.

*Induced oscillatory activity*

We investigated induced oscillatory activity to test if the main effect of cTBS on task-related beta-band activity related to the modulation of non-phase-locked oscillatory responses. As for the total oscillatory beta analyses, the induced beta-band activity analyses revealed that F3, F4 and P3 electrodes showed a significant decrease in beta band power (13-19 Hz, 140-370 ms) in the DLPFC cTBS compared to the vertex cTBS (cluster-corrected permutation test, $P = 0.022$, Supplementary Fig. 1A, B). A time-frequency analysis on significant channels is shown in Supplementary Fig. 1C. Finally, we conducted cluster-based statistical analyses to evaluate the interaction between cTBS sites and contexts in the beta-band (13-19 Hz, 140-370 ms). No significant interaction effect was found in the beta frequency range (cluster-corrected permutation test, $P > 0.10$).

------------------------------------------------------------

Insert Figure 4 around here

------------------------------------------------------------

**Discussion**

Although enhanced cognitive control improves goal-directed behavior, there is evidence that reduced cognitive control is beneficial for automatic behaviors and implicit forms of learning. The DLPFC is thought to exert top-down cognitive control by means of inhibition of automatic (implicit) processes. Thus, DLPFC deactivation should significantly facilitate implicit learning. In the present study, we investigated whether reduced cognitive control through DLPFC disruption could enhance implicit contextual memory-guided attention. In addition, we investigated whether DLPFC disruption and concomitant reduced cognitive control would be reflected in decreased beta-band oscillatory activity. As predicted, we found that cTBS applied to the DLPFC led to a robust increase in implicit contextual memory performance compared with cTBS applied to the vertex as a control condition. Moreover, cTBS over the DLPFC significantly decreased beta-band oscillatory activity (13-19 Hz, 140-370 ms) at fronto-central channels.

*TMS and contextual cueing task*

Our behavioral results were consistent with previous findings (Chun and Jiang 1998; Chun and Phelps 1999; Johnson et al. 2007; Manelis and Reder 2012). In the visual search task, participants were highly accurate and showed a general RT facilitation for repeated contexts in both cTBS conditions. Importantly, as an earlier experiments our results of the recognition task indicated that repetition facilitation effect occurs implicitly (Chun and Jiang 1998; Chun and Jiang 2003).

Our results show that disruption of DLPFC processing using cTBS enhanced implicit contextual memory. This result provided a first link between DLPFC and implicit contextual memory-guided attention. The observed improvement is in line with previous observations that DLPFC disruption by cTBS facilitates implicit recognition (Lee et al. 2013). On a more general level, our finding is consistent with previous findings suggesting that disengagement of top-down cognitive control through DLPFC deactivation exerts a facilitative effect on implicit processes (Limb and Braun 2008; Liu et al. 2012; Amer et al. 2016). Thus, our TMS results support the notion of DLPFC involvement in top-down cognitive control through suppression of automatic (implicit) processes (Kübler et al. 2006; de Manzano et al. 2012), and provide first causal evidence for the interfering role of DLPFC on implicit contextual learning.

In accordance with previous studies investigating the effects of the duration of cTBS on cortical inhibition, we observed that implicit contextual learning was significantly higher for cTBS DLPFC than vertex cTBS during epochs 2-5. Previous observations showed that cTBS delivered over a period of 40 s reduced brain activity few minutes after stimulation and for up to one hour, with a maximum effect at 15 to 40 min after cTBS (Huang et al. 2005). Thus, the enhanced performance we report here may reflect sufficient cortical inhibition since the beginning of epoch 2 (~8 min after cTBS) until the end of the task, suggesting that the acquisition of implicit contextual memory can happen fast, but top-down cognitive control delays the expression of implicit memory when this task is applied in normal conditions.

In line with previous studies on inter-individual variation in implicit contextual learning (Lleras and von Mühlenen 2004; Schlagbauer et al. 2012), our results show that in the cTBS vertex condition 72% of the individuals showed the contextual cueing effect. On the contrary, in the cTBS DLPFC condition 100% of the subjects exhibited the contextual cueing effect. This observation supports the notion that inter-individual variation in contextual cueing performance could be explained by the degree of pressure on the attentional system exerted by voluntary cognitive control. High cognitive control (e. g., an active search strategy) through DLPFC participation competes for attentional resources dedicated for processing repeated contexts, abolishing the contextual cueing effect (Schlagbauer et al. 2012). Similar results were found when a spatial working memory task was combined with the contextual cueing task. In this case, working memory load took away attentional resources reducing implicit contextual learning (Manginelli et al. 2012; Annac et al, 2013). Importantly, working memory processing is supported by DLPFC (Cohen et al. 1997; Courtney et al. 1997; Smith et al. 1998). We assume that top-down cognitive control might narrow the focus of attention on target information, and a main part of contextual information remains excluded of both the encoding and retrieval processes. Therefore, reduced cognitive control through DLPFC disruption may have boosted implicit learning by broadening the scope of attention. This mechanism could not only facilitate implicit memory encoding but also later access. Similar mechanisms have been proposed to explain the facilitatory effect of reduced cognitive control on creativity and problem solving (Amer et al. 2016).

*TMS and beta-band activity*

We found that task-related beta-band activity (13-19 Hz, 140-370 ms) at fronto-central channels was reduced in cTBS DLPFC compared with cTBS vertex. Our finding is in agreement with a previous study that investigated the after-effect of rTMS over DLPFC on oscillatory responses in healthy subjects. Woźniak-Kwaśniewska et al. (2014) observed that low beta-band oscillatory responses (14-22 Hz) at frontal electrodes decreased after cTBS over the DLPFC. Since beta and gamma oscillations are related to activity of fast-spiking inhibitory interneurons (Cardin et al. 2009), decreased beta-band activity could be interpreted as an indicator of reduced cortical inhibition. Thus, the reduced beta-band activity we reported here may reflect reduced top-down signaling mechanism and consequent diminution of the capacity to produce neural ensemble synchrony necessary to enable voluntary cognitive control.

Several studies indicated that beta-band activity plays an important role in top-down cognitive control. Beta-band synchronization between frontal and parietal areas was observed during attentional top-down processing but not during bottom-up processing (Buschman and Miller 2007). In addition, prominent phase synchronization in the beta-band range between frontal and parietal areas was reported during endogenously driven choices in comparison to stimulus driven choices (Pesaran et al. 2008). This suggests that beta-band activity is prominent in settings that involve a strong top-down cognitive control, whereas reduced beta-band activity is more related to tasks that involve exogenous, bottom-up processing (Engel and Fries 2010). In the light of the above findings, we suggest that the decreased beta-band oscillatory power observed here may be indicative of reduced TMS-induced top-down cognitive control. Therefore, the facilitatory effect of TMS DLPFC on implicit contextual learning could be the result of the suppression of top-down, endogenous processing and concomitant speeding of bottom-up stimulus-driven processing, necessary for encoding and recovering implicit memories. Importantly, our results showed that cTBS reduced the beta-band oscillatory power independently of the contexts, suggesting that it exclusively reflects a top-down cognitive control mechanism but not implicit memory processes.

There are some limitations of the current design. First, our study does not allow for the source reconstruction of beta-band activity since the EEG was recorded from 30 scalp electrodes, an insufficient number of electrodes to provide adequate accuracy in source estimation (Sohrabpour et al. 2015; Song et al. 2015). Another potential limitation of this study is that we could not examine if the facilitatory effect of TMS DLPFC on implicit contextual learning occurs in the encoding, retrieval or both phases. Future research would benefit from incorporating this issue into the study design.

In conclusion, our study shows that disruption of DLPFC both significantly improves implicit contextual learning and decreased task-related beta-band oscillatory power. To our knowledge, this is the first causal evidence that the DLPFC plays an interfering role in implicit contextual memory-guided attention in the human brain. Furthermore, our study sheds light onto the relationship between brain oscillations and implicit memory guided attention and supports the notion that beta-band oscillatory activity promotes DLPFC-mediated top-down control. Future research is needed to determine the specific mechanisms by which DLPFC interacts with the brain systems that support implicit contextual memory-guided attention.

**Acknowledgements**

This research was supported by grant SFB636 ⁄ C1 from the Deutsche Forschungsgemeinschaft to HF and Administrative Department of Science, Technology and Innovation (Colciencias) to MR and CAPES/Alexander von Humboldt (AvH) grants to JC. We thank Christopher Milde for his help in the translation of research materials into German.

**Bibliography**

Annac E, Manginelli AA, Pollmann S, Shi Z, Müller HJ, Geyer T. 2013. Memory under pressure: Secondary-task effects on contextual-cueing of visual search. J Vis. 13(13):6.

Amer T, Campbell KL, Hasher L. 2016. Cognitive Control As a Double-Edged Sword. Trends Cogn. Sci. 20:905-292.

Bell A, Sejnowski T. 1995. An information-maximization approach to blind separation and blind deconvolution. Neural Comput. 7:1129-1159.

Brainard DH. (1997). The Psychophysics Toolbox, Spat Vis. 10:433-436.

Buschman TJ, Miller EK. 2007. Top-down versus bottom-up control of attention in the prefrontal and posterior parietal cortices. Science. 315:1860-1862.

Cardin JA, Carlen M, Meletis K, Knoblich U, Zhang F, et al. 2009. Driving fast-spiking cells induces gamma rhythm and controls sensory responses. Nature. 459:663–667.

Cohen JD, Perlstein WM, Braver TS, Nystrom LE, Noll DC, Jonides J, Smith EE. 1997. Temporal dynamics of brain activation during a working memory task. Nature. 386:604-608.

Cole M, Schneider W. 2007. The cognitive control network: integrated cortical regions with dissociable functions. Neuroimage. 37:343-360.

Corbetta M, Shulman GL. 2011. Spatial neglect and attention networks. Annu Rev Neurosci. 34:569-99.

Corbetta M, Shulman GL. 2002. Control of goal-directed and stimulus-driven attention in the brain. Nat Rev Neurosci. 3:201-15.

Corbetta M, Kincade JM, Ollinger JM, McAvoy MP, Shulman GL. 2000. Voluntary orienting is dissociated from target detection in human posterior parietal cortex. Nat Neurosci. 3:92-297.

Courtney SM, Ungerleider LG, Keil K, Haxby JV. 1997. Transient and sustained activity a distributed neural system for human working memory. Nature. 386:608-611.

Chun MM, Jiang Y. 1998. Contextual cueing: implicit learning and memory of visual context guides spatial attention. Cogn Psychol 36: 28-71.

Chun MM. Jiang Y. 2003. Implicit, long-term spatial contextual memory. J. Exp. Psychol. Learn.

Mem. Cognit. 29:224-234.

Chun MM, Phelps EA. 1999. Memory deficits for implicit contextual information in amnesic subjects with hippocampal damage. Nat Neurosci. 2:844-847.

Delorme A, Makeig S. 2004. EEGLAB: an open source toolbox for analysis of single-trial EEG dynamics. J Neurosci Methods. 134:9-21.

de Manzano Ö, Ullén F. 2012. Goal-independent mechanisms for free response generation: creative and pseudo-random performance share neural substrates. Neuroimage. 59:772-780.

Egeth HE, Yantis S. 1997. Visual attention: control, representation, and time course. Annu Rev Psychol. 48:267–97.

Engel AK, Fries P. 2010. Beta-band oscillations-signalling the status quo?. Curr Opin Neurobiol. 20:156–65.

Engel AK, Fries P, Singer W. 2001. Dynamic predictions: oscillations and synchrony in top-down processing. Nat Rev Neurosci. 2:704-716.

Huang YZ, Edwards MJ, Rounis E, Bhatia KP, Rothwell JC. 2005. Theta burst stimulation of the human motor cortex. Neuron. 45:201–6.

Hutchinson JB, Turk-Browne NB. 2012. Memory-guided attention: control from multiple memory systems. Trends Cogn Sci. 16:576-579.

Iversen JR, Repp BH, Patel AD. 2009. Top-down control of rhythm perception modulates early auditory responses. Ann N Y Acad Sci. 1169:58-73.

Johnson JS, Woodman GF, Braun E, Luck SJ. 2007. Implicit memory influences the allocation of attention in visual cortex. Psychon Bull Rev. 14:834-839.

Kristjánsson A. Campana G. 2010. Where perception meets memory: a review of repetition priming in visual search tasks. Atten Percept Psychophys. 72:5–18.

Kübler A, Dixon V, Garavan H. 2006. Automaticity and reestablishment of executive control-an fMRI study. J Cogn Neurosci. 18:1331–42.

Lee TG, Blumenfeld RS, D'Esposito M. 2013. Disruption of dorsolateral but not ventrolateral prefrontal cortex improves unconscious perceptual memories. J Neurosci. 33:13233-13237.

Limb CJ, Braun AR. 2008. Neural substrates of spontaneous musical performance: an FMRI study of jazz improvisation. PLoS ONE. 3:e1679.

Liu S, Chow HM, Xu Y, Erkkinen MG, Swett KE, Eagle MW, Rizik-Baer DA, Braun AR. 2012. Neural correlates of lyrical improvisation: an fMRI study of freestyle rap. Sci Rep. 2:1-8.

Lleras A, Von Mühlenen A. 2004. Spatial context and top-down strategies in visual search. Spat Vis. 17:465–482.

Manelis A, Reder LM. 2012. Procedural learning and associative memory mechanisms contribute to contextual cueing: Evidence from fMRI and eye-tracking. Learn Mem. 9:527-534.

Manginelli AA, Geringswald F, Pollmann S. 2012. Visual search facilitation in repeated displays depends on visuospatial working memory. J Exp Psychol. 59:47-54.

Maris E, Oostenveld R. 2007. Nonparametric statistical testing of EEG- and MEG-data. J Neurosci Methods. 164:177–190.

Mitra PP, Pesaran B. 1999. Analysis of dynamic brain imaging data. Biophys J. 76:691–708.

Moores E, Laiti L, Chelazzi L. 2003. Associative knowledge controls deployment of visual selective attention. Nat Neurosci 6:182–189.

Okazaki M, Kaneko Y, Yumoto M, Arima K. 2008. Perceptual change in response to a bistable picture increases neuromagnetic beta-band activities. Neurosci Res. 61:319-328.

Oldfield R. 1971. The assessment and analysis of handedness: The Edinburgh inventory. Neuropsychologia. 9:97–113.

Oostenveld R, Fries P, Maris E, Schoffelen JM. 2011. FieldTrip: Open source software for advanced analysis of MEG, EEG, and invasive electrophysiological data. Comput Intell Neurosci. 2011:156869.

Pesaran B, Nelson MJ, Andersen RA. 2008. Free choice activates a decision circuit between frontal and parietal cortex. Nature. 453:406-409.

Rossi S, Hallett M. et al. 2009. Safety, ethical considerations, and application guidelines for the use of transcranial magnetic stimulation in clinical practice and research. Clin Neurophysiol. 120:2008–2039.

Schlagbauer B, Müller HJ, Zehetleitner M, Geyer T. 2012. Awareness in contextual cueing of visual

search as measured with concurrent access- and phenomenal-consciousness tasks. J Vis. 12:1-12.

Shulman GL, Ollinger JM, Akbudak E, Conturo TE, Snyder AZ, Petersen SE, Corbetta M. 1999. Areas involved in encoding and applying directional expectations to moving objects. J. Neurosci. 19:9480-9496.

Slepian D. 1978. Prolate spheroidal wave functions, Fourier analysis, and uncertainty—V: the discrete case. Bell Syst Tech J. 57:1371-1430.

Smith EE, Jonides J, Marshuetz C, Koeppe RA. 1998. Components of verbal working memory: evidence from neuroimaging. Proc Natl Acad Sci USA. 95:876-882.

Sohrabpour A, Lu Y, Kankirawatana P, Blount J, Kim H, He B. 2015. Effect of EEG electrode number on epileptic source localization in pediatric patients Clin Neurophysiol. 126:472-480.

Song J, Davey C, Poulsen C, et al. 2015. EEG source localization: sensor density and head surface coverage. J Neurosci Methods. 256:9–21.

Staudigl T, Hanslmayr S. 2013. Theta oscillations at encoding mediate the context-dependent nature of human episodic memory. Curr Biol. 23:1101-1106.

Stokes MG, Atherton K, Patai EZ, Nobre AC. 2012. Long-term memory prepares neural activity for perception. Proc Natl Acad Sci USA. 109:E360-367.

Waldhauser GT, Braun V, Hanslmayr S. 2016. Episodic memory retrieval functionally relies on very rapid reactivation of sensory information. J Neuro. 36:251-260.

Wassermann EM. 1996. Risk and safety of repetitive transcranial magnetic stimulation: report and suggested guidelines from the International Workshop on the Safety of Repetitive Transcranial Magnetic Stimulation. Electroencephalogr Clin Neurophysiol. 108:1-16.

Wozniak-Kwasniewska A, Szekely D, Aussedat P, Bougerol T, David O. 2014. Changes of oscillatory brain activity induced by repetitive transcranial magnetic stimulation of the dorsolateral prefrontal cortex in healthy subjects. Neuroimage. 88:91-99.

Zhao J, Al-Aidroos N, Turk-Browne NB. 2013. Attention is spontaneously biased toward regularities. Psychological Science. 24(5):667-677.

**Figure legends**

**Figure 1.** Behavioral Task. Example of a block of the contextual cueing task. Each participant performed 20 blocks separated by 10 sec. Each block contained 24 randomly interleaved trials of 12 repeated and 12 new contexts. Each trial began with a central cross presented for 0.5 sec followed by a search array which was presented for a maximum of 3 sec or until a response was made. Participants were required to indicate by pressing keys on a keyboard whether the target letter T was rotated 90° to the left or 90° to the right. A variable duration ITI (0.5 - 1 sec) separated subsequent trials.

**Figure 2.** Definition of the left dorsolateral prefrontal cortex (DLPFC) used as TMS target. The DLPFC target was defined by the coordinates (-43, 35, 30) in MNI152 space and transformed into each participant's native space for frameless stereotaxy. The arrows indicate the location of the DLPFC target shown on a template brain in MNI152 space.

**Figure 3.** Behavioral performance in the contextual cueing task. A) Mean reaction times (with standard error of the mean) for cTBS DLPFC (top) or cTBS vertex (bottom) as a function of epoch and context. B) The contextual-cueing effect (with standard error of the mean) as a function of cTBS site and epoch.

**Figure 4.** Analysis of total oscillatory activity. (A) Results from the two-step time-frequency analysis. Time-frequency window (13-19 Hz, 140-370 ms, dashed box) for DLPFC versus vertex contrast but not for the repeated versus new context contrast was identified by sliding window analyses, and survived a cluster-based permutation test. On the right, the topography of the results from the cluster-based dependent t-test randomization procedure over the window of interest (13-19 Hz, 140-370 ms) for the DLPFC versus vertex contrast is depicted (pcorr < 0.05). Channels showing a significant interaction are highlighted. (B) Topographic distribution of significant post-stimulus beta-band activity (13-19 Hz, 140-370 ms). Topographies are depicted for cTBS DLPFC, cTBS vertex, and their differential activity (cTBS DLPFC versus cTBS vertex). (C) Time-frequency

representations of total oscillatory activity at F3, F4 and P3 channels. Time-frequency representations are depicted for cTBS DLPFC, cTBS vertex, and their differential activity (i.e., cTBS DLPFC versus cTBS vertex). Vertical dashed lines indicate stimulus onset.

**Supplementary Figure 1.** Analysis of the induced oscillatory activity. (A) The topography of the results from the cluster-based dependent t-test randomization procedure over the window of interest (13-19 Hz, 140-370 ms) for the cTBS DLPFC versus cTBS vertex contrast is depicted (pcorr < 0.05). Channels showing a significant interaction are highlighted. (B) Topographic distribution of significant post-stimulus beta-band activity (13-19 Hz, 140-370 ms). Topographies are depicted for cTBS DLPFC, cTBS vertex, and their differential activity (cTBS DLPFC versus cTBS vertex). (C) Time-frequency representations of induced oscillatory activity at F3, F4 and P3 channels. Time-frequency representations are depicted for cTBS DLPFC, cTBS vertex, and their differential activity (cTBS DLPFC versus cTBS vertex). Vertical dashed lines indicate stimulus onset.